\documentclass[fleqn]{2023SCGE}
\usepackage[normalem]{ulem}
\usepackage[retainorgcmds]{IEEEtrantools}

\newcommand{\LNO}{La$_4$Ni$_3$O$_{10}$}
\newcommand{\XO}{$d_{x^2-y^2}$~}
\newcommand{\ZO}{$d_{z^2}$~}

\usepackage{cancel}

\begin{document}

\ensubject{subject}

\ArticleType{Article}
\SpecialTopic{SPECIAL TOPIC: }
\Year{}
\Month{}
\Vol{}
\No{}
\DOI{}
\ArtNo{}
\ReceiveDate{xxx}
\AcceptDate{xxx}
\OnlineDate{xxx}

\title{Magnetically Mediated Cross-Layer Pairing in Pressurized Trilayer Nickelate \LNO}
{Magnetically Mediated Cross-Layer Pairing in Pressurized Trilayer Nickelate \LNO}

\author[1,2]{Jialin Chen}{}%
\author[1,3]{Chuanshu Xu}{}%
\author[1,4]{Qiaoyi Li}{}%
\author[1,2]{Wei Li}{{w.li@itp.ac.cn}}

\AuthorMark{J. Chen}

\AuthorCitation{Jialin Chen, Chuanshu Xu, Qiaoyi Li, and Wei Li}

\address[1]{Institute of Theoretical Physics, Chinese Academy of Sciences, Beijing 100190, China}
\address[2]{Hefei National Laboratory, Hefei 230088, China}
\address[3]{School of Physics, Xiamen University, Xiamen 361005, China}
\address[4]{School of Physical Sciences, University of Chinese Academy of Sciences, Beijing 100049, China}

\abstract{The recently discovered trilayer nickelate superconductor \LNO~under pressure has emerged as a promising platform for exploring unconventional superconductivity. However, the pairing mechanism remains a subject of active investigations. With large-scale density matrix renormalization group calculations on a realistic two-orbital trilayer Hubbard model, we elucidate the superconducting (SC) mechanism in this system. Our results reveal distinct magnetic correlations in the two different orbitals: while the \ZO orbital exhibits both interlayer and cross-layer antiferromagnetic (AFM) correlations, the \XO orbital shows exclusively cross-layer AFM correlations, rendering a quasi-long-range SC order in the latter. We demonstrate that the Hund's rule coupling is essential for forming the SC order, and discuss the effects of kinetic AFM correlation and Hubbard repulsive $U$. Our findings motivate a further simplification of the trilayer Hubbard to an effective bilayer mixed-dimensional Hubbard model, providing a unified framework for understanding interlayer SC in both trilayer and bilayer nickelates.
}


\PACS{ }
\maketitle
\begin{multicols}{2}

\section{Introduction}\label{sec:intro}

The recent discovery of high-temperature superconductivity (SC) in pressurized La$_3$Ni$_2$O$_7$, with a critical temperature $T_c \simeq 80$~K~\cite{Nickelate80K}, has stimulated considerable research interest in the Ruddlesden-Popper (RP) perovskite series R$_{n+1}$Ni$_n$O$_{3n+1}$ (R= La, Pr, Nd). Shortly thereafter, high-$T_c$ SC has also been identified in the $n=3$ member, \LNO, also under high pressure and with $T_c \approx 20-30$~K~\cite{Zhu2024Superconductivity, Zhang2025Superconductivity, Li2024Signature, Kakoi2024Multiband, Sakakibara2024Theoretical, Li2024Structural} and about $40$~K in Pr$_4$Ni$_3$O$_{10}$~\cite{Zhang2025Bulk, pei2024pressure, Huang2024Signature, Pei2025Unveiling}. These experimental advances have established nickelates with RP phases as a cornucopia for realizing high-temperature SC states. \Authorfootnote The experimental findings also pose fundamental questions regarding (1) the pairing mechanism, (2) the dominant orbital character of the SC order parameter, (3) its layer-number ($n$) dependence, etc. Despite extensive theoretical studies on La$_3$Ni$_2$O$_7$~\cite{Luo2023Model, zhang2023electronic, yang2023possible, lechermann2023electronic, gu2023effective, shen2023effective, christiansson2023correlated, Shilenko2023Correlated, wu2023charge, cao2023flat, chen2023critical, liu2023spmwave, lu2023interlayer, zhang2023structural, oh2023type, liao2023electron, qu2023bilayer, yang2023minimal, jiang2023high, zhang2023trends, huang2023impurity, qin2023hightc, tian2023correlation, lu2023superconductivity, jiang2023pressure, kitamine2023theoretical, luo2023hightc, zhang2023strong, pan2023effect, sakakibara2023possible, lange2023pairing, geisler2023structural, yang2023strong, rhodes2023structural, lange2023feshbach, labollita2023electronic, kumar2023softening, kaneko2023pair, lu2023interplay, ryee2023critical, schlomer2023superconductivity, liu2024evolution, Zhan2025Cooperation, Yang2024Strong, Xia2025Sensitive, Ouyang2024Hund, Sui2024Electronic, Zheng2025s, Botzel2024Theory, Wang2024Electronic, Geisler2024Optical, Heier2024Competing, chang2023fermi,  abadi2024electronic, Yi2024Nature, wu2024ac3ni2o7, Lechermann2024Electronic, Ouyang2024Absence, KK2024Oxygens, Jiang2025Intertwined, Chen2025Oxygen, Huo2025Modulation} and \LNO~\cite{sakakibara2023possible, Lu2025Superconductivity, Leonov2024Electronic, Chen2024Trilayer, Yang2024Effective, Zhang2025SDW, Zhang2024Prediction, Huang2024Interlayer, Oh2025TypeII, Qin2024Frustrated, huo2024electron, Yang2024Decomposition, LaBollita2024Electronic, Wang2024NFL, Tian2024Effective}, these fundamental questions remain subjects of intense debate in the field.

For example, in the RP nickelate with $n=2$ there is a critical open question and ongoing debate regarding orbital-selective pairing mechanisms --- with some studies favoring the \XO~orbital dominance~\cite{lu2023interlayer, Chen2024Orbital, oh2023type, liao2023electron, qu2023bilayer, lechermann2023electronic, jiang2023pressure, Fan2024Superconductivity, qu2025hund, lu2023superconductivity, pan2023effect,lange2023pairing,lange2023feshbach} while others advocate for \ZO orbital primacy~\cite{Liu2023s, yang2023minimal, qin2023hightc, yang2023possible, sakakibara2023possible, luo2023hightc, Ryee2024Quenched, shen2023effective, kaneko2023pair, gu2023effective, Zhan2025Cooperation}.  
For \LNO, theoretical approaches including mean-field theory (MF)~\cite{Lu2025Superconductivity,Huang2024Interlayer}, random-phase approximation (RPA)~\cite{Zhang2024SC,Zhang2024Prediction}, functional renormalization group (FRG)~\cite{Yang2024Effective}, and static auxiliary-field quantum Monte Carlo (SAFQMC)~\cite{Qin2024Frustrated} identify the \ZO orbital as the primary host of SC pairing. However, this scenario appears to conflict with experimental observations~\cite{gim2025orbital,shi2025absence}.

In particular, recent experimental studies present compelling challenges to theoretical models of SC in \LNO. Intriguingly, the absence of SC in I4/mmm-AP samples --- regardless of the presence or absence of \ZO-derived Fermi surface pockets~\cite{shi2025absence} --- questions the proposed \ZO orbital theory in high-pressure SC phase. Complementary electronic Raman scattering data further highlight the predominant involvement of \XO orbitals in density wave formation \cite{gim2025orbital}, while observed magnetic structures align with an AFM spin density wave (SDW) localized on outer Ni layers \cite{Zhang2020Intertwined, khasanov2025identical}. These findings collectively emphasize the essential role of \XO-orbital in forming the SC order. Owing to the current scarcity of high-precision, large-scale many-body numerical investigations, systematic studies are urgently required to resolve this outstanding question regarding the  SC pairing mechanism in \LNO.

In this work, we investigate the SC pairing mechanism for the RP nickelate \LNO~through large-scale density matrix renormalization group (DMRG) calculations on a realistic two-orbital trilayer Hubbard model incorporating both $e_g$ orbitals (\XO~and \ZO) of the Ni atoms. Using realistic parameters, we find that the \XO~orbital dominates the $s$-wave SC pairing in \LNO, while the \ZO~orbital contributes only to SC fluctuations. The quasi-long-range SC order of \XO~orbital is due to cross-layer pairing between the two outer layers of trilayer system, which requires the existence of a Hund's coupling ($J_\mathrm{H} \gtrsim 0.5$\,eV) and is mediated by antiferromagnetic (AFM) correlation. The origin of these AFM correlations and the influence of the Hubbard 
$U$ parameter are discussed, along with predictions for further enhancing the SC order. Based on these findings, we propose an effective mixed-dimensional (mixD) bilayer Hubbard Hamiltonian~\cite{qu2023bilayer} as a single-orbital minimal model capturing the SC pairing in \LNO, putting it in the same unified framework of SC pairing with the $n=2$ RP nickelate superconductor La$_3$Ni$_2$O$_7$.

\section{Model and method}  
\subsection{Two-orbital trilayer Hubbard model}{\label{Sec:model_2orb_Hub}}   
The proposed real-space pairing scenarios in \LNO~include: (i) cross-layer pairing between the two outer layers \cite{Yang2024Effective, Zhang2024SC}; (ii) interlayer pairing between inner and outer layers \cite{Lu2025Superconductivity, Qin2024Frustrated, Huang2024Interlayer}; and (iii) intralayer pairing within either the outer layer \cite{Lu2025Superconductivity} or the inner layer \cite{Yang2024Effective}. To address this question and determine the dominant orbital (\XO or \ZO), we consider a two-orbital trilayer Hubbard model for \LNO~\cite{Georges2013Hund, Chen2024Trilayer}, whose Hamiltonian 
$H = H_t + H_I$ contains
$H_t$ and $H_I$ representing the kinetic and interaction terms, respectively.

\begin{figure}[H]
\centering
\includegraphics[width=0.9\linewidth]{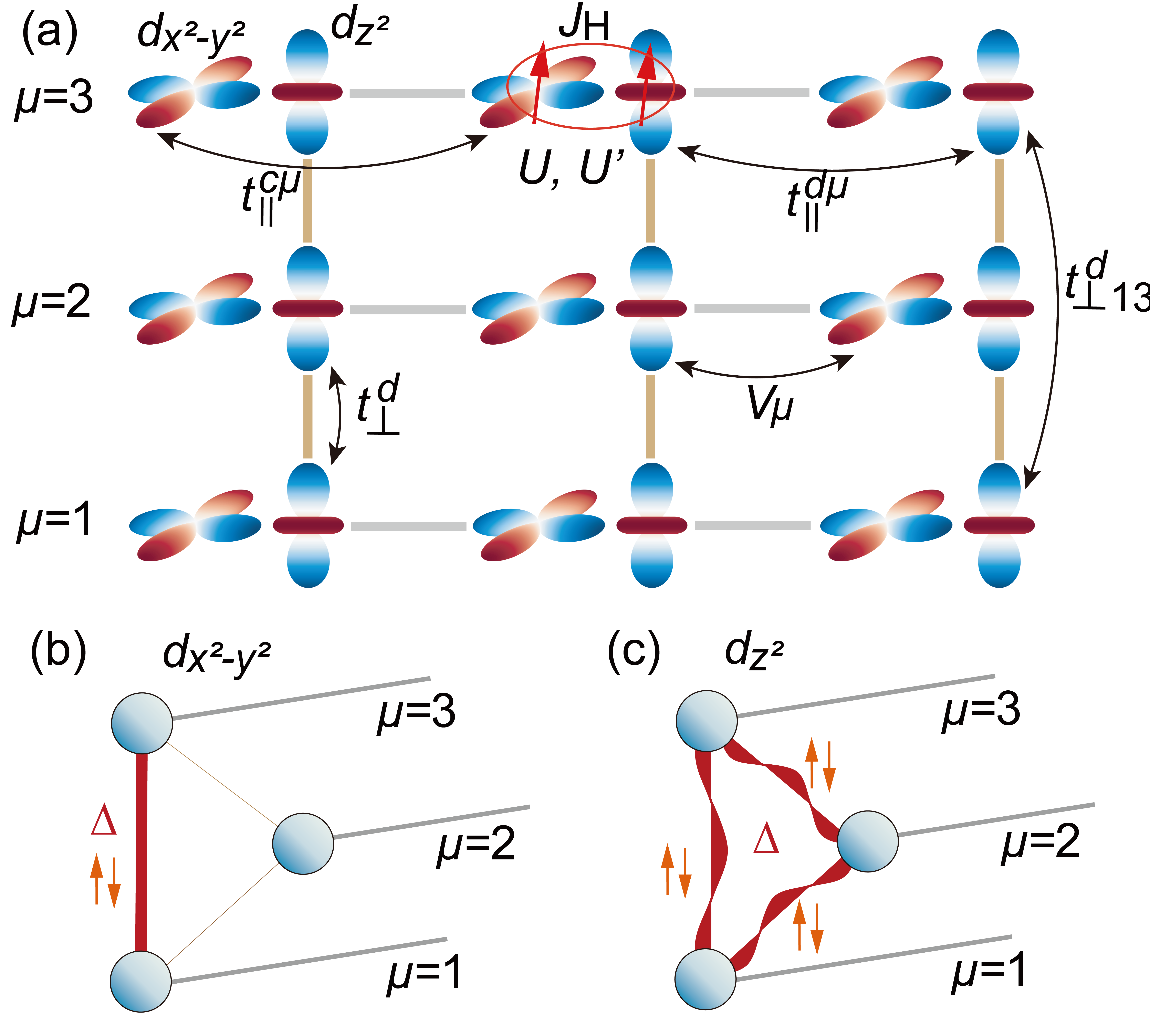} 
\caption{(a) The two-orbital trilayer Hubbard model for \LNO. The dominant hopping and interaction terms are shown, with $\mu=1,2,3$ labeling the layer number. The \XO orbital has significant intra-layer hopping $t^{c\mu}_\parallel$, but its inter-layer hopping is negligible. The \ZO orbital has strong $t^d_\perp$ between the inner ($\mu=2$) and outer layers ($\mu=1,3$), and weak intra-layer ($t^{d\mu}_\parallel$) and cross-layer ($t^d_{\perp 13}$) hopping, while other hopping terms are negligibly small. The parameters $U$, $U'$, and $J_\mathrm{H}$ denote the intra-orbital Coulomb repulsion, inter-orbital Coulomb repulsion, and Hund's coupling, respectively. Diagrammatic illustration of (b) \XO and (c) \ZO orbitals showing inter-layer and cross-layer AFM spin correlations (orange up-down arrows) and superconducting pairing ($\Delta$). The \ZO orbital exhibits both spin and pairing frustration stemming from its triangular geometry between the three layers, whereas the \XO orbital largely evades such frustration owing to its spatially non-uniform electron distribution.
}
\label{Fig1}
\end{figure}

As illustrated in Fig.~\ref{Fig1}(a), the kinetic terms (including chemical potentials) are as follows,
\begin{equation}
	\begin{split}
		&H_t = - \sum_{\langle i, j\rangle, \mu, \sigma} V_{\mu} \left(c_{i,\mu, \sigma}^{\dagger} d_{j, \mu, \sigma} + d_{i,\mu, \sigma}^{\dagger} c_{j, \mu, \sigma} + \mathrm{H.c.}\right)  \\
        &-\sum_{\langle i, j\rangle, \mu, \sigma} t^{c\mu}_{\parallel} \left(c_{i, \mu, \sigma}^{\dagger} c_{j, \mu, \sigma} + \mathrm{H.c.}\right)         
		-\sum_{i, \sigma; \mu=1,3} t^{c}_{\perp} \left(c_{i, 2, \sigma}^{\dagger} c_{i, \mu, \sigma} + \mathrm{H.c.}\right) \\  
        &-\sum_{\langle i, j\rangle, \mu, \sigma} t^{d\mu}_{\parallel} \left(d_{i, \mu, \sigma}^{\dagger} d_{j, \mu, \sigma} + \mathrm{H.c.}\right)       
		-\sum_{i, \sigma; \mu=1,3} t^{d}_{\perp} \left(d_{i, 2, \sigma}^{\dagger} d_{i, \mu, \sigma} + \mathrm{H.c.}\right) \\   
		&-\sum_{i,\sigma} t^{c}_{\perp 13} \left(c_{i, 1, \sigma}^{\dagger} c_{i, 3, \sigma} + \mathrm{H.c.}\right)	          
		-\sum_{i,\sigma} t^{d}_{\perp 13} \left(d_{i, 1, \sigma}^{\dagger} d_{i, 3, \sigma} + \mathrm{H.c.}\right) \\
        &+ \sum_{i, \mu} \left(\varepsilon^{c}_{\mu} n_{i, \mu}^{c} 
        + \varepsilon^{d}_{\mu} n_{i, \mu}^{d}\right), \\
		\label{Eq:2orb-hop}
	\end{split}
\end{equation}  
where $c_{i, \mu, \sigma}$ ($d_{i, \mu, \sigma}$) denotes the annihilation operator for an electron in the \XO (\ZO) orbital at site $i$, layer $\mu \in \{1,2,3\}$, with spin $\sigma \in \{\uparrow, \downarrow\}$. Here, $n_{i,\mu}^{\alpha}$ denotes the density operator at site $i$ and layer $\mu$, for $\alpha=c$ or $\alpha=d$, corresponding to \XO or \ZO orbital. The parameters are defined as follows: (1) $t^{\alpha\mu}_\parallel$, the intralayer hopping for orbital $\alpha$ in layer $\mu$; (2) $t^{\alpha}_\perp$, the interlayer hopping between the inner layer ($\mu=2$) and outer layers ($\mu=1,3$); (3) $t^{\alpha}_{\perp 13}$, the cross-layer hopping between the two outer layers ($\mu=1$ and $\mu=3$); (4) $V_\mu$, the inter-orbital hybridization between \XO and \ZO orbitals on the nearest-neighboring sites $i,j$ and in the same layer $\mu$; (5) $\varepsilon^{\alpha}_{\mu}$, the on-site energy for orbital $\alpha$ in layer $\mu$.  

We adopt the hopping parameters derived from density functional theory (DFT) calculations~\cite{Chen2024Trilayer}. The specific parameter values in the inner layer are $t^{c2}_\parallel = 0.521$ eV, $t^{d2}_\parallel = 0.168$ eV, $V_2 = -0.298$ eV, $\varepsilon^c_2=1.094$ eV, $\varepsilon^d_2=1.081$ eV, in the out layers $t^{c1}_\parallel = t^{c3}_\parallel= 0.511$ eV,  $t^{d1}_\parallel = t^{d3}_\parallel = 0.143$~eV, $V_1 = V_3 = -0.274$~eV. The interlayer hoppings are $t^d_{\perp} = 0.738$ eV and  $t^c_{\perp} = 0.033$ eV, and the cross-layer hoppings are $t^d_{\perp 13} = 0.078$ eV and $t^c_{\perp 13} = 0$. The chemical potential values for the two orbitals are $\varepsilon^c_1=\varepsilon^c_3=0.867$~eV, $\varepsilon^d_1=\varepsilon^d_3=0.683$~eV. The non-interacting band for the two-orbital trilayer model in two dimensions is shown in Appendix~\ref{AppSec:2dband}. Furthermore, we fix an average filling of $n = 2/3$ electrons per orbital per site in our calculations, corresponding to the nominal $3d^{7.33}$ configuration of Ni atoms in \LNO.

For the interaction terms, we consider the Kanamori Hamiltonian \cite{Georges2013Hund}:
\begin{equation}
	\begin{split}
		H_I = &~U\sum_{i,\mu,\alpha} n^{\alpha}_{i,\mu,\uparrow} n^{\alpha}_{i,\mu,\downarrow} 
		+ U'\sum_{i,\mu,\sigma} n^{c}_{i,\mu,\sigma} n^{d}_{i,\mu,\bar{\sigma}} \\ 
		& + (U'-J_\mathrm{H})\sum_{i,\mu,\sigma} n^{c}_{i,\mu,\sigma} n^{d}_{i,\mu,\sigma} \\
		& + J_\mathrm{H} \sum_{i,\mu} \big( c^{\dagger}_{i,\mu,\uparrow} 
		c^{\dagger}_{i,\mu,\downarrow} d_{i,\mu,\downarrow} d_{i,\mu,\uparrow} \\
		&- c^{\dagger}_{i,\mu,\uparrow} 
		c_{i,\mu,\downarrow} d^{\dagger}_{i,\mu,\downarrow} d_{i,\mu,\uparrow} + \mathrm{H.c.} \big),
		\label{Eq:U-term}
	\end{split}
\end{equation}
with $U$, $U'=U-2J_\mathrm{H}$, and $J_\mathrm{H}$ the intra-orbital Coulomb repulsion, inter-orbital Coulomb repulsion, and Hund's rule coupling, respectively. For \LNO, we take typical parameter values as $J_\mathrm{H} \approx 1$ eV and $U\approx3.5$ eV \cite{du2025dichotomy}. 

\subsection{Method}
 In this study, we investigate the ground-state properties using the DMRG method \cite{White1992Density, SCHOLLWOCK2011The}. 
DMRG is a variational tensor-network method based on the matrix product state representation of the many-body wavefunction \cite{Klumper1991, Fannes1992, Klumper1993}. By systematically truncating the Hilbert space according to the entanglement spectrum, it achieves controllable accuracy with polynomial computational cost. These properties make DMRG one of the most powerful tools available for studying strongly correlated quantum systems in low dimensions.

The DMRG method has several advantages over other many-body techniques such as dynamical mean-field theory (DMFT)~\cite{Georges1996DMFT}, RPA~\cite{Pines1952RPA,Scalapino2012A}, FRG~\cite{Metzner2012FRG}, and SAFQMC~\cite{Mayr20005Phase, Dubi2007Nature}. First, DMRG provides numerically unbiased and highly accurate results for strongly correlated systems, particularly in one- and quasi-one-dimensional geometries. Unlike perturbative approaches such as RPA and FRG, which rely on weak- or intermediate-coupling expansions, DMRG captures strong correlation effects without assuming a small parameter. In contrast to DMFT, which treats local correlations exactly but approximates nonlocal ones, DMRG reliably includes both local and nonlocal correlations. Moreover, unlike SAFQMC, which neglects quantum fluctuations along the imaginary-time axis, DMRG fully accounts for quantum correlations. These advantages make it especially suitable for investigating unconventional superconductivity, spin and charge ordering, and topological phases.

Our DMRG simulations employ a lattice of length $L=48$ and width $W=3$, where we implement both Abelian charge conservation and non-Abelian spin symmetries to enhance computational efficiency using TensorKit.jl~\footnote{https://github.com/Jutho/TensorKit.jl} and  FiniteMPS.jl~\footnote{https://github.com/Qiaoyi-Li/FiniteMPS.jl}. Additionally, we incorporate controlled bond expansion techniques \cite{Gleis2023PRLControlled} to speed up the calculations while maintaining numerical accuracy. The DMRG calculations are performed with a maximum bond dimension $D^*$ of 9096 multiplets, equivalent to about $D=25000 ~U(1)$ states, achieving truncation errors $\epsilon \lesssim 3\times10^{-5}$. To ensure reliable measurements of correlation functions, we evaluate correlation functions within the central half of the lattice and perform systematic bond-dimension extrapolations to mitigate finite-bond-dimension effects.

\section{Results}
\subsection{Charge distributions and local spin correlations}  

\begin{figure}[H]
	\includegraphics[width=1\linewidth]{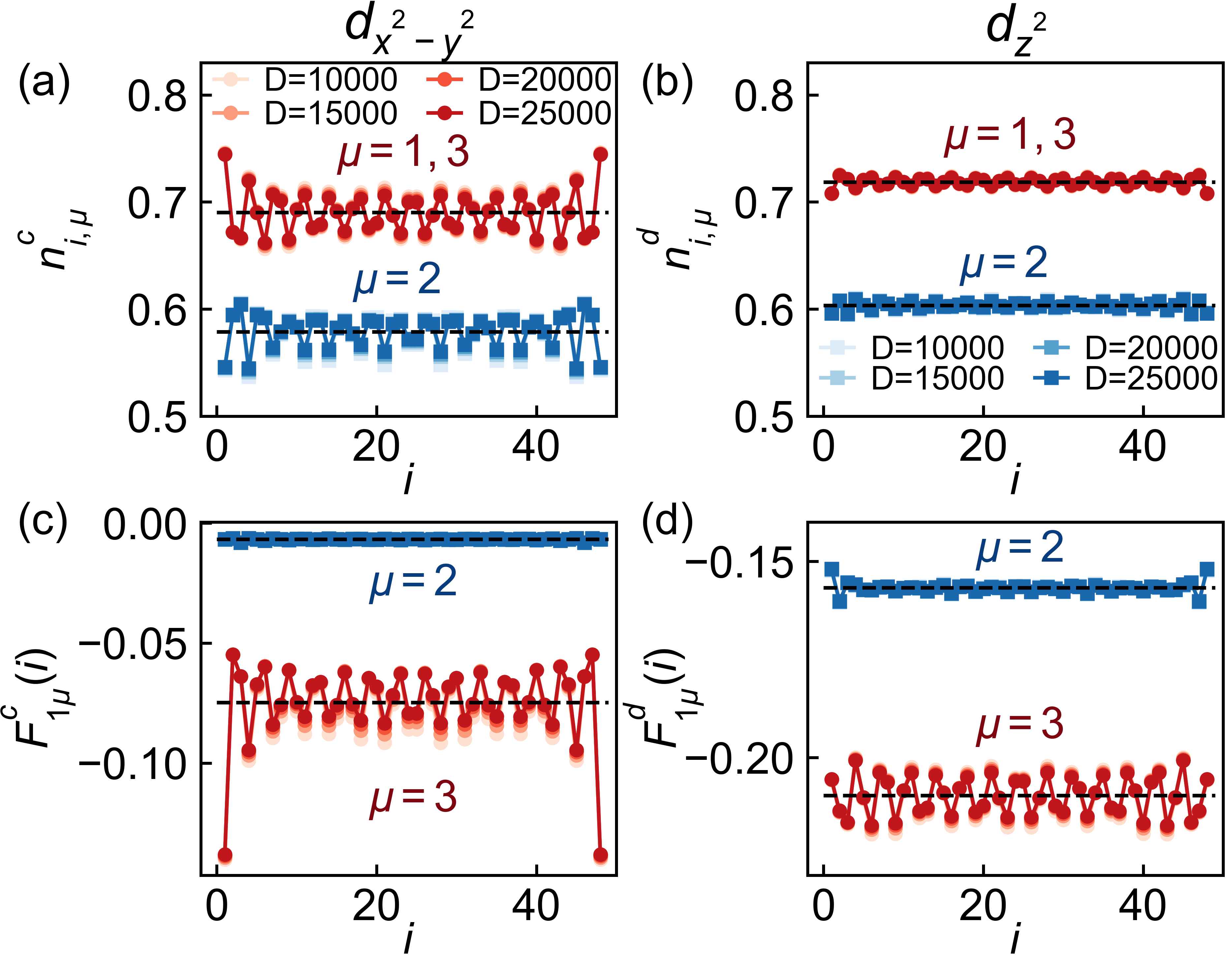} 
	\caption{Local properties of the two-orbital trilayer Hubbard model with $U=3.5$ eV and $J_\mathrm{H}=1.0$ eV. The charge density distribution of the (a) \XO and (b) \ZO orbitals in the outer layer (red) and inner layer (blue). The interlayer (blue) and cross-layer (red) spin correlation for the (c) \XO and (d) \ZO orbitals. Average values are shown with back dashed lines. Results obtained with bond dimensions $D=10000,15000,20000, 25000$ are displayed using progressively darker colors.}
	\label{Fig2}
\end{figure}

We consider the two-orbital Hubbard model with representative parameters $U = 3.5$~eV and $J_\mathrm{H} = 1$~eV. Figures~\ref{Fig2}(a) and (b) display the orbital-resolved charge distributions $n^\alpha_{i,\mu}$ for the \XO and \ZO orbitals, respectively. In both orbitals, the outermost atomic layer exhibits a higher average charge density compared to the inner layers. A comparative analysis reveals that the \ZO orbital possesses a marginally higher charge density than the \XO orbital across both inner and outer layers. Notably, the \XO orbital exhibits a pronounced CDW modulation characterized by a wave vector $k = 0.58\pi$ in both inner and outer layers. While significantly weaker, a CDW feature with the same wave vector is also observed in the \ZO orbital, potentially induced via the hybridization.

\begin{figure*} 
\includegraphics[width=1\linewidth]{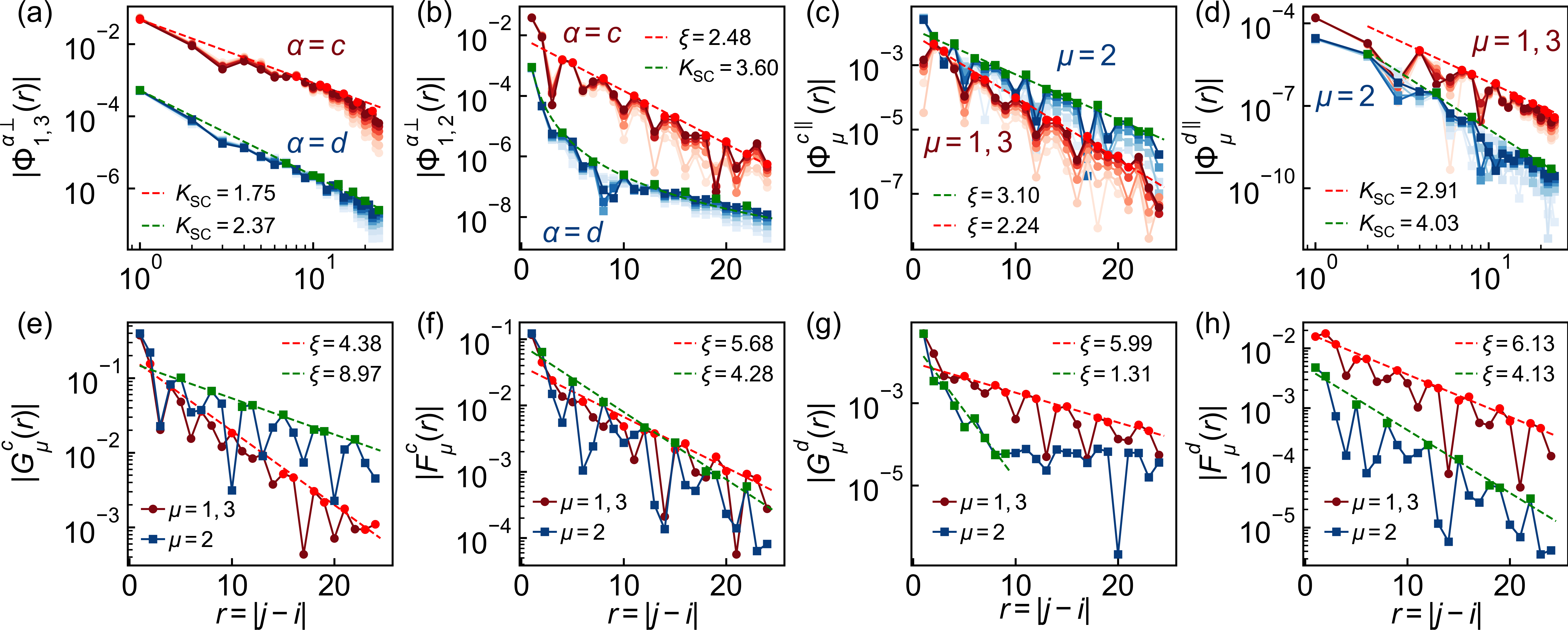} 
\caption{Correlation functions for the two-orbital trilayer Hubbard model with $U=3.5$ eV and $J_\mathrm{H}=1.0$ eV. (a) Cross-layer pairing correlation $\Phi^{\alpha\perp}_{1,3}(r)$ and (b) interlayer pairing correlation $\Phi^{\alpha\perp}_{1,2}(r)$ for \XO (red) and \ZO (blue) orbitals. Intralayer pairing correlations for (c) \XO and (d) \ZO orbitals, comparing outer (red) and inner (blue) layers. The SC correlation functions are presented for progressively increasing bond dimensions $D = 8000$, $10000$, $15000$, $20000$, $25000$ and the extrapolated $D \rightarrow \infty$ limit, with each curve depicted in sequentially darker colors. (e, g) Single-particle Green's functions and (f,h) intralayer spin correlation functions for the \XO and \ZO orbitals, showing only $D \rightarrow \infty$ extrapolated data (inner layer: blue; outer layer: red).
The data involved in the fittings are marked in bright red or green, i.e., in the same color code as the fitting line.}
\label{Fig3}
\end{figure*}

Figures~\ref{Fig2}(c) and (d) display the interlayer spin correlation $F^{\alpha}_{12}(i)=\langle \mathbf{S}^\alpha_{i,1}\cdot \mathbf{S}^\alpha_{i,2}\rangle$ and cross-layer spin correlation $F^{\alpha}_{13}(i) = \langle \mathbf{S}^\alpha_{i,1}\cdot \mathbf{S}^\alpha_{i,3}\rangle$ for both \ZO and \XO orbitals. The \ZO orbital shows AFM correlations for both interlayer and cross-layer couplings [Fig.~\ref{Fig2}(d)], with the cross-layer correlation being slightly stronger. As illustrated in Fig.~\ref{Fig1}(c), these AFM correlations create geometric frustration in the three-layer spin configuration, leading to competing SC pairing channels as they share the limited available electrons. We will systematically analyze the origin of these AFM correlations in Section~\ref{Sec_ThreeSites}.

In striking contrast, the \XO orbital exhibits negligible direct interlayer and cross-layer hopping, yet develops AFM correlations mediated through ferromagnetic Hund's coupling with the \ZO orbital \cite{lu2023interlayer,oh2023type,qu2025hund,Tian2024Correlation,Chen2024Orbital,qu2023bilayer,Lu2025Superconductivity}. The cross-layer AFM spin correlation in the \XO orbital reaches $-0.077$ [Fig.~\ref{Fig2}(c)], an order of magnitude larger than its interlayer correlation ($-0.007$). This behavior differs fundamentally from the \ZO orbital, where spin correlations remain localized between outer layers without inducing frustration, and no pairing frustration occurs [Fig.~\ref{Fig1}(b)]. These computational results align well with experimental observations of AFM coupling between outer Ni layers \cite{Zhang2020Intertwined, khasanov2025identical}.

\subsection{Quasi-long-range cross-layer SC pairing}  
To investigate SC correlations, we define a spin-singlet SC pairing operator between site $i$ in layer $\mu$ and site $j$ in layer $\nu$ for orbital $\alpha$ as:
\begin{equation}
	\Delta^{\alpha\dagger}_{i,\mu, j,\nu } = \frac{1}{\sqrt{2}} \left( c^{\alpha, \dagger}_{i,\mu,\uparrow} c^{\alpha, \dagger}_{j,\nu,\downarrow} - c^{\alpha, \dagger}_{i,\mu,\downarrow} c^{\alpha, \dagger}_{j,\nu,\uparrow} \right).
\end{equation}
The interlayer (cross-layer) SC correlation function between layer $\mu=1$ and $\mu=2$ ($\mu=3$) is given by $\Phi^{\alpha\perp}_{1,\mu} (r) = \langle \Delta^{\alpha\dagger}_{i,1, i,\mu} \Delta^{\alpha}_{j,1, j,\mu} \rangle$ where $r = |j-i|$, while the intralayer correlation function on layer $\mu$ is $\Phi^{\alpha\parallel}_{\mu} (r) = \langle \Delta^{\alpha\dagger}_{i,\mu, i+1,\mu} \Delta^{\alpha}_{j,\mu, j+1,\mu} \rangle$. These correlations were computed with bond dimensions $D=8000$-$25000$, followed by extrapolation to the $D\rightarrow\infty$ limit to extract the Luttinger parameter $K_{sc}$ (for power-law decay $\Phi(r) \sim r^{-K_\mathrm{SC}}$) or correlation length $\xi$ (for exponential decay $\Phi(r) \sim e^{-r/\xi}$).

Figures~\ref{Fig3}(a-d) present the SC correlation functions for both $e_g$ orbitals, with fixed $U = 3.5$ eV and $J_\mathrm{H} = 1$ eV. The cross-layer correlation $\Phi^{c\perp}_{1,3} (r)$ [panel (a)] exhibits significantly greater magnitude than all other pairing channels, with a fitted Luttinger parameter $K_\mathrm{SC}=1.75$ -- the only instance below the critical value of 2. This establishes cross-layer pairing ($s$-wave pairing) in the \XO orbital as the sole quasi-long-range ordered SC channel.

While other pairing channels lack quasi-long-range order, they exhibit substantial SC fluctuations. The \XO orbital shows exponentially decaying correlations for both $\Phi^{c\perp}_{1,2}(r)$ and $\Phi^{c\parallel}_\mu (r)$. In contrast, all \ZO orbital correlations follow power-law decay, with distinct Luttinger parameters: $K_\mathrm{SC}=$ 2.36 ($\Phi^{d\perp}_{1,3}$), 3.61 ($\Phi^{d\perp}_{1,2}$), 2.91 ($\Phi^{d\parallel}_{1}$, $\Phi^{d\parallel}_{3}$), and 4.02 ($\Phi^{d\parallel}_{2}$). This analysis reveals that dominant SC fluctuations originate from the \ZO orbital, particularly in cross-layer pairing and intralayer pairing within the outer layers.

Figure~\ref{Fig3}(e) displays the extrapolated single-particle Green's functions $G^{\alpha}_{\mu}(r) = \sum_{\sigma}\langle c^{\alpha\dagger}_{i,\mu,\sigma}c^{\alpha}_{j,\mu,\sigma}\rangle$ (with $r=|j-i|$) for the \XO orbitals, revealing exponential decay ($G^c_1(r)\sim e^{-r/\xi}$) with correlation lengths $\xi=4.38$ (outer layer) and $\xi=8.97$ (inner layer), indicating finite single-particle gaps. The corresponding intralayer spin correlations $F^{\alpha}_{\mu}(r) = \langle\mathbf{S}^{\alpha}_{i,\mu}\cdot\mathbf{S}^{\alpha}_{j,\mu}\rangle$ in Fig.~\ref{Fig3}(f) similarly show exponential decay. The gapped character of both charge and spin excitations in the \XO orbital's outer layers provide further confirmation of the quasi-long-range SC order observed in its cross-layer channel.  
 
For comparison, Fig.~\ref{Fig3}(g) displays the single-particle Green's functions $G^d_\mu(r)$ of the \ZO orbital. The outer layers ($G^d_1$ and $G^d_3$) exhibit exponential decay $G^d_1(r)\sim e^{-r/\xi}$ with correlation length $\xi=5.98$, while the inner layer shows rapid initial decay ($\xi=1.31$) followed by a saturation-like behavior at larger distances. Corresponding spin correlations $F^d_\mu(r)$ in panel (h) demonstrate exponentially decaying behavior for both layers, confirming gapped spin excitations throughout the system. Notably, the outer layer exhibits significantly stronger spin correlations --- approximately tenfold greater in magnitude and with longer correlation lengths—than the inner layer, underscoring the amplified spin correlation effects within the outer regions of the \ZO orbital.

\subsection{Hund's rule coupling and SC pairing}
  
As the Hund's coupling $J_\mathrm{H}$ may vary in nickelates \cite{Pardo2010Quantum, Kumar2025Softening, du2025dichotomy}, we systematically examines the SC pairing in the two-orbital Hubbard model by varying $J_\mathrm{H}$ . Within the relevant parameter range, the cross-layer pairing $\Phi^{c\perp}_{1,3}$ emerges as the only channel exhibiting quasi-long-range SC order, and the Hund's coupling plays an important role for the cross-layer pairing.

\begin{figure}[H]
\includegraphics[width=1\linewidth]{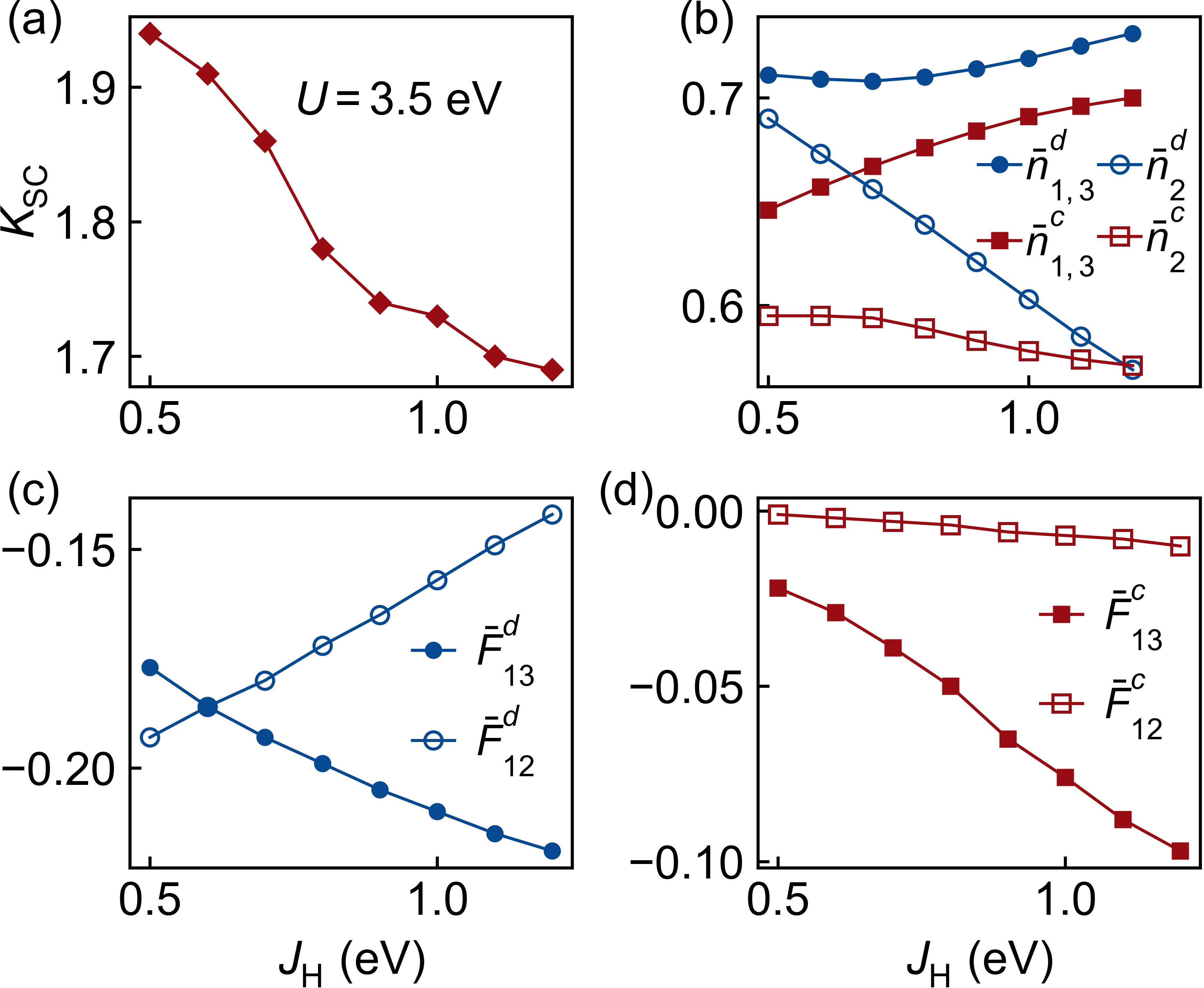} 
\caption{ Impact of Hund's coupling $J_\mathrm{H}$ on the two-orbital trilayer Hubbard model with fixed $U=3.5$ eV. (a) Luttinger parameter $K_\mathrm{SC}$ for cross-layer pairing $\Phi^{c\perp}_{1,3}$. (b) Average charge densities $\bar{n}^\alpha_\mu$ for both \ZO (circle) and \XO (square) orbitals. (c-d) Average interlayer and cross-layer spin correlation for (c) \ZO and (d) \XO orbitals. All presented data correspond to the $D\to\infty$ limit.
}
\label{Fig4}
\end{figure}

Figure~\ref{Fig4} systematically investigates the influence of Hund's coupling ($J_\mathrm{H}\in[0.5,1.2]$ eV) in the realistic model at fixed $U=3.5$~eV. Panel (b) reveals that $J_\mathrm{H}$ strongly modulates the layer-dependent charge distribution: increasing $J_\mathrm{H}$ enhances (reduces) the charge density in outer (inner) layers for both orbitals, with the most pronounced variation occurring in $\bar{n}^d_2$. This charge distributions consequently influence the spin correlations, as revealed in panel (c), where the \ZO orbital's cross-layer (interlayer) spin correlations strengthen (weaken) with increasing $J_\mathrm{H}$.

The enhanced FM Hund's coupling enables more efficient transfer of AFM interactions from the \ZO to \XO orbital. Combined with the charge redistribution effect, this produces a dramatic fivefold enhancement of the \XO orbital's cross-layer spin correlation [see Fig.~\ref{Fig4}(d)], across the studied $J_\mathrm{H}$ range.  
Notably, the interlayer spin correlation remains negligibly small due to competing effects between decreasing charge density and strengthening orbital bounding. 
Consequently, the Luttinger parameter $K_\mathrm{SC}$ for cross-layer pairing $\Phi^{c\perp}_{1,3}$ systematically decreases with increasing $J_\mathrm{H}$ [Fig.~\ref{Fig4}(a)], revealing a significantly strengthened SC order. The emergence of quasi-long-range SC order for $J_\mathrm{H} \gtrsim 0.5$~eV underscores the key role of Hund's rule coupling in stabilizing the cross-layer SC in \LNO.

\subsection{Effect of Coulomb interaction}{\label{sec:Coulomb}} 
  
\begin{figure}[H]
	\includegraphics[width=1\linewidth]{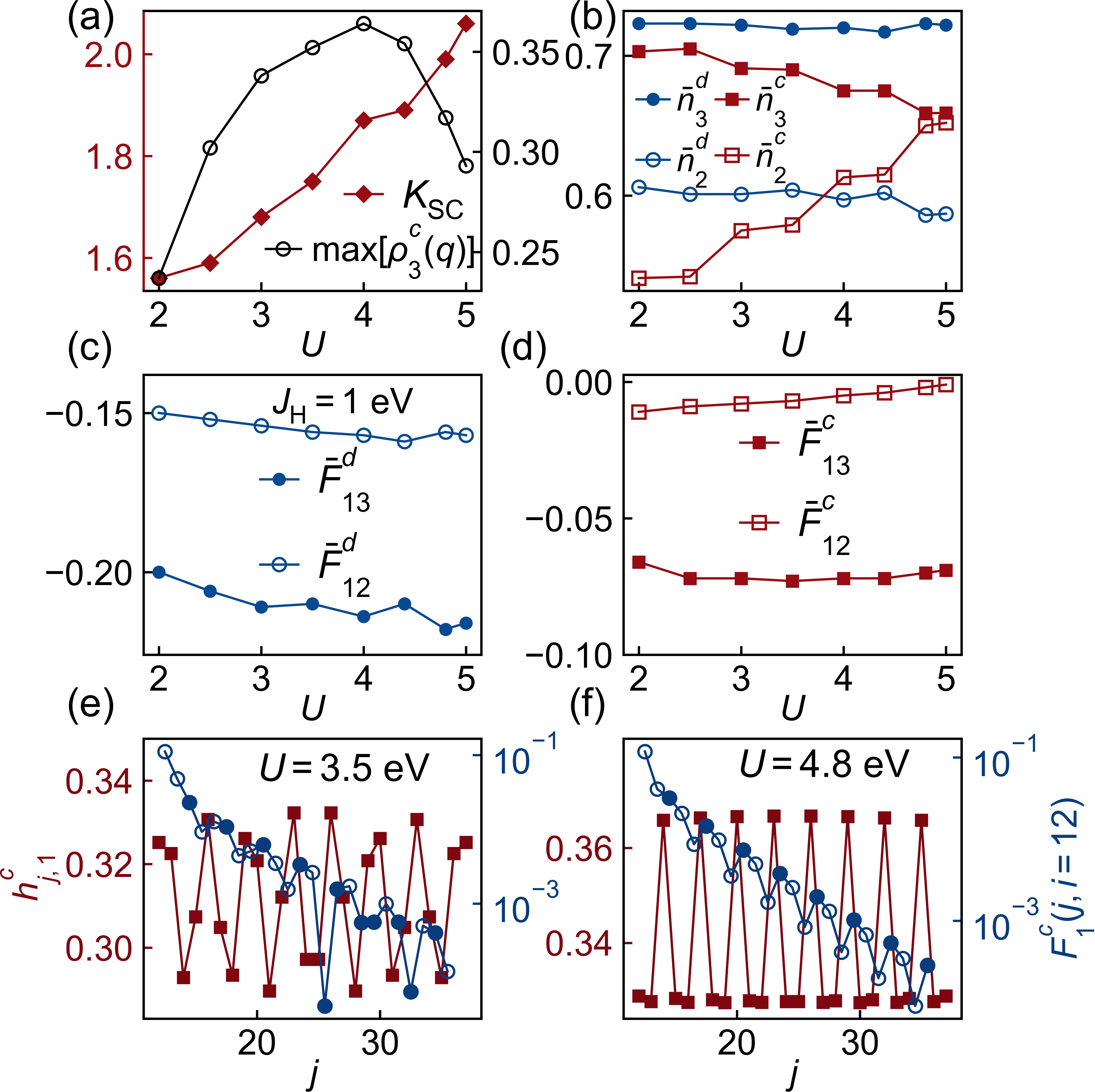} 
	\caption{  Impact of Coulomb interaction strength $U$ on the trilayer Hub-
bard mode with fixed Hund's coupling $J_\mathrm{H}=1.0$ eV. (a) Luttinger parameter $K_\mathrm{SC}$ for cross-layer pairing $\Phi^{c\perp}_{1,3}$ (red) and the maximum value of $\rho^c_3(q)$ (black) in the outer layers of \XO orbital. (b) Orbital-resolved average charge densities $\bar{n}^\alpha_\mu$ for both \ZO (circle) and \XO (square) orbitals. (c-d) Average interlayer and cross-layer spin correlation for (c) \ZO and (d) \XO orbitals.  Hole density $h^c_{j,1}=1-n^c_{j,1}$ and spin correlation $F^c_1(j,i=12)$ in the outer layer (e.g., $\mu=1$) of the \XO orbital for (e) $U=3.5$ eV and (f) $U=4.8$ eV. Positive (negative) values of $F^c_1(j,i=12)$ are marked with filled (hollow) circles.  All presented data correspond to the $D\to\infty$ limit. }
	\label{fig_U}
\end{figure}

Figure~\ref{fig_U} examines the dependence on Coulomb interaction strength ($U \in [2,5]$ eV) in the trilayer two-orbital Hubbard model with fixed $J_\mathrm{H}=1$ eV. The cross-layer spin correlation $\langle\mathbf{S}^d_{i,1}\cdot\mathbf{S}^d_{i,3}\rangle$ increases gradually with $U$ [panel (c)], whereas the interlayer correlation $\langle\mathbf{S}^d_{i,1}\cdot\mathbf{S}^d_{i,2}\rangle$ remains nearly independent of $U$. Clear orbital differentiation emerges in the charge response: the \ZO orbital shows little redistribution, while the \XO orbital undergoes pronounced layer-dependent charge transfer, with $\bar{n}^c_3$ decreasing and $\bar{n}^c_2$ increasing as $U$ grows [panel (b)]. Owing to the competing effects of enhanced $\langle \mathbf{S}^d_{i,1}\cdot\mathbf{S}^d_{i,3}\rangle$ and reduced $\bar{n}^c_3$, the cross-layer spin correlation in the \XO orbital remains strikingly robust [panel (d)]. These results indicate that AFM correlations are not the dominant factor governing the evolution of SC with $U$. 

 We further probe possible CDW order in the outer layers of the \XO orbital by evaluating the modulation amplitude of the charge density,
\begin{equation}
\rho^c_\mu(q) = \Big|\sum_{i} e^{iq r_i}\big(\langle n^c_{i,\mu}\rangle - \bar{n}^c_{\mu}\big)\Big|.
\end{equation}
As shown in Fig.\ref{fig_U}(a), the CDW amplitude grows markedly with $U$ in the range $[2,4]$ eV, introducing strong competition with superconductivity. At stronger coupling (e.g., $U=4.8$ eV), the system undergoes a transition to a stripe-ordered ground state [Fig.\ref{fig_U}(c)], characterized by a well-defined period-three modulation of the outer-layer hole density $h^c_{j,1}=1-n^c_{j,1}$. Remarkably, the modulation peaks align precisely with the sign-changing nodes of the intralayer spin correlation function $F^c_{1}(j,i=12)$, referenced to site $i=12$. Consequently, as shown in panel (a), $K_\mathrm{SC}$ for $\Phi^{c\perp}_{1,3}$ increases with $U$, exceeding the critical threshold of $2$ for $U \gtrsim 4.5$ eV. This behavior signals the suppression of SC at strong Coulomb coupling due to emergent charge order. Taken together, our results for \LNO~demonstrate that intermediate interaction strengths ($U \in [2,4]$ eV) are optimal for stabilizing SC, consistent with experimental evidence of moderate electronic correlations in \LNO~\cite{liu2024evolution}.  

\section{Discussion}   
\subsection{Kinetic AFM correlations} \label{Sec_ThreeSites} 
To elucidate the origin of cross-layer AFM correlations in \LNO, we analyze their emergence in the \ZO orbital through a minimal, three-site single-orbital Hubbard model:

\begin{equation}{\label{Eq:threesites}}
\begin{split}		
	H =  -\sum_{\sigma, \mu\in\{1,3\}} t^{d}_{\perp} \left(d_{2, \sigma}^{\dagger} d_{ \mu, \sigma}  
	+ \mathrm{H.c.}\right)  
	+ \sum^3_{\mu=1} U n_{\mu}^{d} 
	+\Delta\varepsilon^{d}_{2} n_{2}^{d},
\end{split}
\end{equation}  
where $\Delta \varepsilon^d_2 \equiv \varepsilon^d_2 - \varepsilon^d_1$ denotes the chemical potential difference between inner and outer layers. Except for the minimized geometry, the hopping parameters remain the same with the two-orbital trilayer model [see Eq.~(\ref{Eq:2orb-hop})].
 
\begin{figure}[H]
\includegraphics[width=0.95\linewidth]{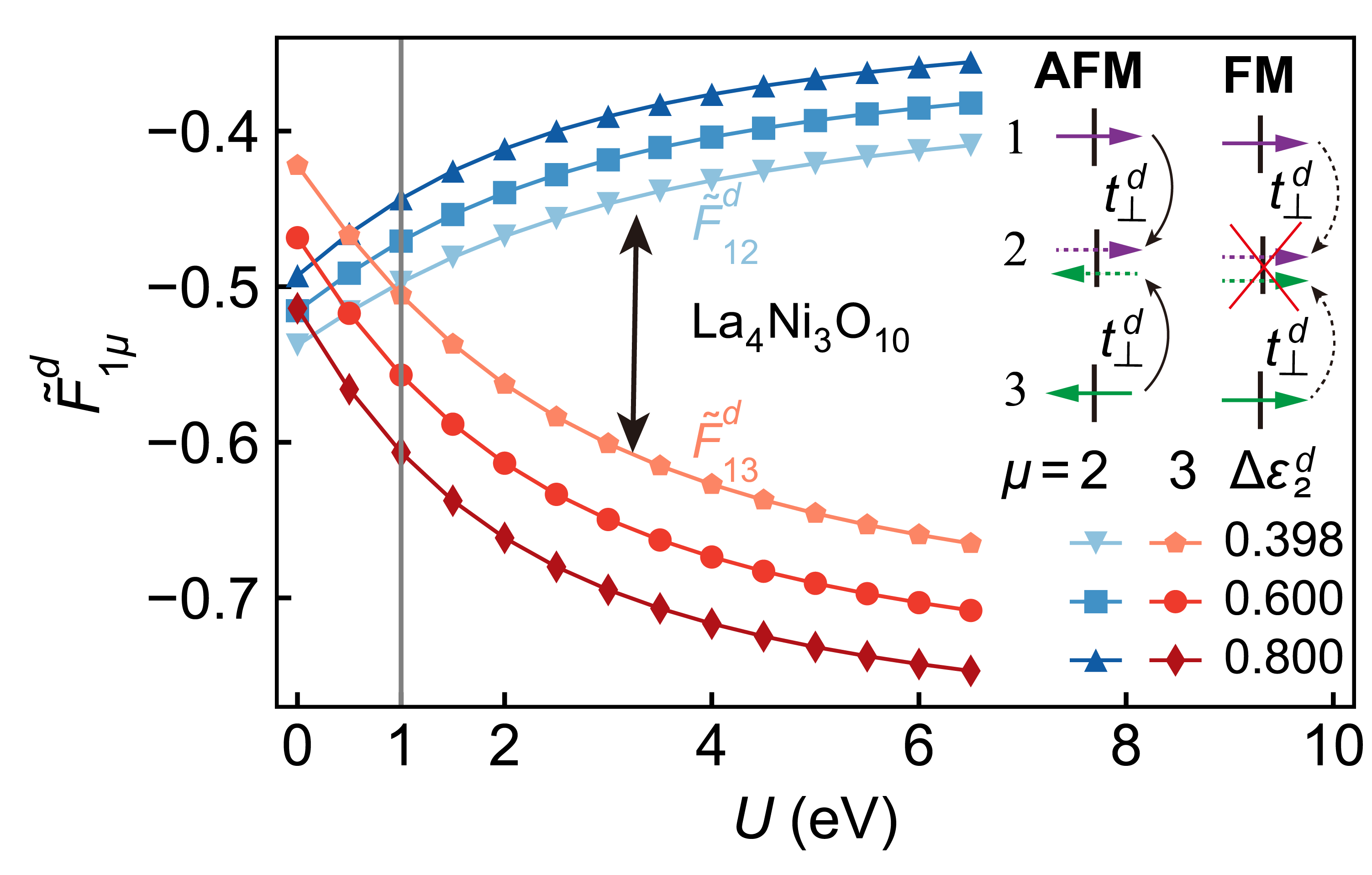} 
\caption{The interlayer (blue) and cross-layer (red) spin correlations for the three-site single-orbital (\ZO) Hubbard model~(\ref{Eq:threesites}). The schematic inset illustrates representative hopping processes for two distinct spin configurations: the AFM channel allows hopping processes, and the FM channel prohibits certain hopping due to the Pauli exclusion principle. 
}
\label{Fig6}
\end{figure}

As illustrated in the inset of Fig.~\ref{Fig6}, in the AFM channel where the total spin projection is $\langle S^z\rangle=0$, hopping processes are permitted. Conversely, these processes are forbidden in the FM channel ($\langle S^z\rangle=1$) by the Pauli exclusion principle. Because allowed hopping reduces the system's kinetic energy, the constraints in the FM channel create an energetically unfavorable state relative to the AFM configuration. This fundamental energetic difference drives the system to spontaneously develop kinetic AFM correlations, which persist even in the non-interacting limit ($U=0$).

To systematically investigate the effects of kinetic AFM correlations and Hubbard $U$, we analyze the normalized spin correlations for the \ZO orbital $\tilde{F}_{1\mu}=\langle \mathbf{S}^d_1 \cdot \mathbf{S}^d_\mu\rangle/\sqrt{\langle(\mathbf{S}^{d}_1)^2\rangle \cdot \langle(\mathbf{S}^{d}_\mu)^2\rangle}$ and show the results in Fig.~\ref{Fig6}.
At $\Delta\varepsilon^d_2=0.398$ eV (realistic parameter for \LNO), both nearest-neighbor (NN, $\tilde{F}^d_{12}$) and next-nearest-neighbor (NNN, $\tilde{F}^d_{13}$) spin correlations show AFM character even at $U=0$, confirming their purely kinetic origin. While $\tilde{F}^d_{12}$ correlations dominate at $U=0$, this hierarchy reverses for $U>1.0$ eV: $\tilde{F}^d_{13}$ ($\tilde{F}^d_{12}$) correlations grow (drop) rapidly and monotonically with $U$. The relation further demonstrates that while the $\tilde{F}^d_{12}$ correlation decreases with increasing $U$, $\tilde{F}^d_{13}$ exhibits a fundamentally different behavior across the entire $U$ range, being enhanced by stronger Coulomb interactions.

Moreover, the layer potential difference $\Delta\varepsilon^d_2$ provides additional control over these AFM correlations. As shown in Fig.~\ref{Fig6}, increasing $\Delta\varepsilon^d_2$ systematically strengthens $\tilde{F}^d_{13}$ correlations while simultaneously weakening $\tilde{F}^d_{12}$ correlations. Lastly, while a small direct hopping $t^d_{\perp 13}$ exists between the outer layers of the \ZO orbital, our analysis in Appendix~\ref{AppSec:t13} reveals that its primary effect is to enhance SC pairing through suppression of CDW order in the \XO orbital's outer layers, rather than significantly contributing to cross-layer AFM correlations.

In summary, our model calculations suggest several strategies for enhancing the SC order in the two-orbital trilayer model and possibly also in the \LNO~compound. Specifically, we identify three factors to strengthen interlayer SC pairing: (i) stronger Hund's rule coupling; (ii) larger chemical potential difference between the inner and outer layers --- akin to amplifying their charge density imbalance; and (iii) suppressed CDW orders in the outer layers.

\subsection{Effective mixed-dimensional model} 
Our comprehensive analysis [Figs.~\ref{Fig2}(d) and \ref{Fig6}] reveals that while the \ZO orbital develops both interlayer and cross-layer spin correlations capable of mediating superconducting pairing [Figs.~\ref{Fig3}(a,b)], these channels are strongly suppressed by two competing effects: (i)  pairing frustration from the limited shared electron density in the three layers [see Fig.~\ref{Fig1}(c)], and (ii) Pauli blocking induced by large interlayer hopping \cite{Hilker2023pairing, qu2023bilayer, Chen2024Orbital}.

Remarkably, the \XO orbital exhibits qualitatively different behavior, maintaining negligible interlayer but substantial cross-layer spin correlations [Fig.~\ref{Fig2}(c)]. This unique magnetic environment enables robust quasi-long-range SC order through cross-layer pairing. The essential physics of this pairing channel can be captured by an effective mixD single-orbital Hubbard model describing the two outer \XO layers:

\begin{equation}
	\begin{split}
		H = & -\sum_{\langle i, j\rangle, \sigma} \sum_{\mu\in\{1,3\}} t^{c\mu}_{\parallel}  \left(c_{i, \mu, \sigma}^{\dagger} c_{j, \mu, \sigma}+ \mathrm{H.c.}\right)  \\
	  &+ \sum_{i, \mu\in\{1,3\}} U n_{i, \mu, \uparrow}^{c} n_{i, \mu, \downarrow}^{c}  
		 +\sum_{i} J^{c*}_{\perp 13}  ~\mathbf{S}_{i,\mu=1}^{c} \cdot \mathbf{S}_{i, \mu=3}^{c} \\  
		\label{Eq:1orb-mixD_Hubbard}
	\end{split}
\end{equation}  
where the effective AFM coupling $J^{c*}_{\perp 13}$ between outer layers originates from the \ZO orbital via Hund's coupling, with a reasonable value corresponding to $J_\mathrm{H}\in[0.5,1.2]$ eV. This framework reveals that the cross-layer pairing in \LNO~shares fundamental similarities with the mechanism proposed for bilayer La$_3$Ni$_2$O$_7$ \cite{lu2023interlayer, Chen2024Orbital, qu2023bilayer, oh2023type, qu2025hund}, particularly in its mediation through interorbital magnetic interactions. This mechanism fundamentally differs from naive attempts to generalize the bilayer mixD $t$-$J$ model of La$_3$Ni$_2$O$_7$ to trilayer systems, where such generalizations have stripe-ordered ground states rather than superconductivity~\cite{Bourgund2025, Schlomer2023Robust}.

Interestingly, a very recent RIXS and Raman spectroscopy study~\cite{zhang2025distinct} reports that a strong on-site Hund’s coupling $J_\mathrm{H}\sim 1$~eV, mediating magnetic interactions between the \ZO\ and \XO\ orbitals, is essential to account for the observed bimagnon excitation. They identify an interlayer coupling in the \ZO\ orbital of about $J_z \sim 112$~meV, and further show that the \XO\ orbital is more itinerant while the \ZO\ orbital is more localized. This is consistent with our picture, in which superconductivity is dominated by the \XO\ orbital, with a cross-layer antiferromagnetic interaction ($\sim 112$~meV, assuming comparable interlayer and cross-layer spin interactions) transferred from the \ZO\ orbital through Hund’s coupling.
Furthermore, a recent ARPES experiment~\cite{auyeung2025universal} comparing bilayer and trilayer nickelates reported that a minimal bilayer model successfully captures the essential fermiology and low-energy band structure in both systems, thereby supporting our proposed effective model.

\section{Summary}
In summary, we have conducted large-scale DMRG calculations on a trilayer two-orbital Hubbard model for pressurized \LNO~using realistic parameters. Our key findings reveal: (1) a pronounced CDW in the \XO orbital with a weaker counterpart in the \ZO orbital;  (2) Coulomb and potential difference enhanced kinetic cross-layer AFM correlations in \ZO; (3) Hund's-mediated cross-layer AFM correlations in \XO orbital; (4) the exclusive emergence of quasi-long-range SC order in the cross-layer pairing channel $\Phi^{c,\perp}_{1,3}(r)$ of the \XO orbital; and (5) the \ZO orbital contributes primarily to SC fluctuations in the $\Phi^{d,\perp}_{1,3}(r)$ and $\Phi^{d\parallel}_1(r)$ channels, limited by frustration between pairing channels and Pauli blocking effects \cite{Hilker2023pairing, qu2023bilayer, Chen2024Orbital}.

Our systematic investigation of parameter dependencies demonstrates the critical role of the Hund's coupling ($J_\mathrm{H} \gtrsim 0.5$ eV) in enabling cross-layer pairing through two complementary mechanisms: (i) mediating the transfer of cross-layer AFM correlations from the \ZO to  the\XO orbital, and (ii) enhancing the outer layer charge density in the \XO orbital. 
 
These insights motivate our proposal of an effective bilayer single-orbital mixD Hubbard model, which not only captures SC pairing in trilayer \LNO~but also offers a unified theoretical framework for understanding RP nickelates across both bilayer~\cite{lu2023interlayer, Chen2024Orbital, qu2023bilayer, oh2023type, qu2025hund} and trilayer systems. Within this framework, the lower SC critical temperature in the trilayer system primarily originates from a weaker effective cross-layer (or interlayer) AFM interaction compared to the bilayer system~\cite{Chen2024Electronic, auyeung2025universal}, a mechanism that we will elaborate on in a forthcoming work.

A natural extension of this work would involve testing the proposed mechanism and mixD model for RP nickelates with higher NiO$_2$ layer counts ($n >3$)~\cite{Pan2022Superconductivity, FerencSegedin2023Limits} or hybrid stacking configurations~\cite{Li2024Design}, such as the 1313~\cite{Chen2024Polymorphism, lu2023interlayer, Wang2024Long, Luo2023Model}, 1212~\cite{shi2025superconductivity}, and 2323~\cite{zhang2024magnetic} structures. 
Since our DMRG calculations are performed on a quasi-one-dimensional geometry, intralayer two-dimensional pairing symmetries such as $d$-wave cannot be fully captured.
Further investigations could explore the model's predictions in both two-dimensional limits~\cite{Chen2022Continuous, Chen2024Orbital, qu2023bilayer} and at finite temperatures~\cite{qu2023bilayer}, which are crucial for elucidating the complete SC properties and addressing unresolved questions—particularly the layer-dependent behavior of $T_c$ as a function of the NiO$_2$ layer count.

\Acknowledgements{Jialin Chen, Chuanshu Xu, Qiaoyi Li, and Wei Li are indebted to Xingzhou Qu and Yubo Liu for stimulating discussions. This work was supported by the National Natural Science Foundation of China (Grant Nos.~12222412, 12047503), National Key Projects for Research and Development of China with Grant No.~2024YFA1409200, Innovation Program for Quantum Science and 
	Technology (Nos.~2021ZD0301900), CAS Project for Young Scientists in Basic Research (Grant No.~YSBR-057), 
	and the Postdoctoral Fellowship Program of CPSF (Grant No.~GZB20240772). We thank the HPC-ITP for
	the technical support and generous allocation of CPU time.  
}

\InterestConflict{The authors declare that they have no conflict of interest.}



\bibliographystyle{scpma}
\bibliography{nickelate}

\begin{appendix}




\setcounter{figure}{0} 
\renewcommand{\thesection}{Appendix}

\section{}
\subsection{Non-interacting band} \label{AppSec:2dband} 
\begin{figure}[H]
\centering
\includegraphics[width=1\linewidth]{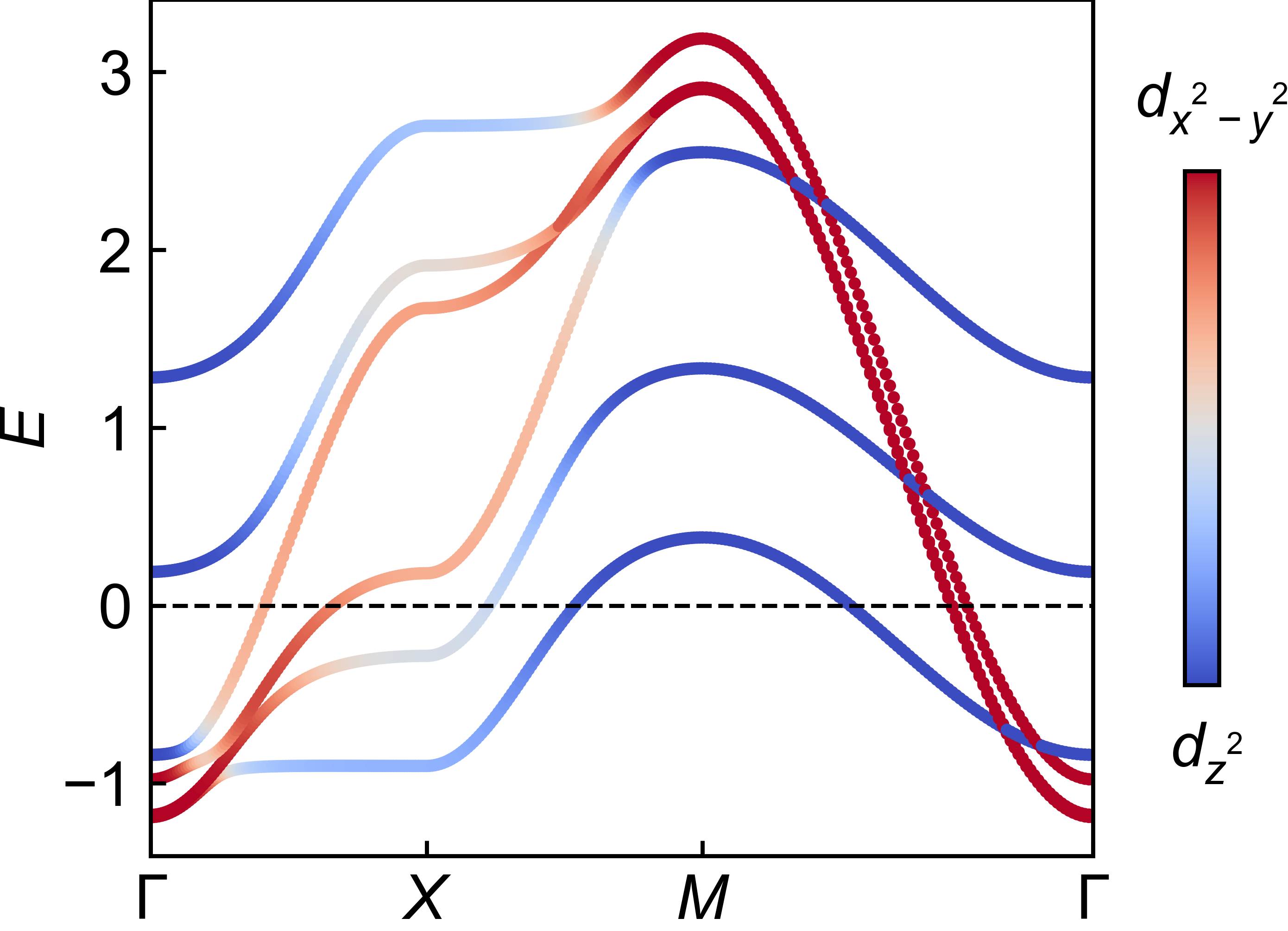} 
\caption{Non-interacting band for the two-orbital trilayer model in two dimensions.
}
\label{Fig_2D_band}
\end{figure} 

In Fig.~\ref{Fig_2D_band}, we present the non-interacting band structure of the two-orbital trilayer model in the two-dimensional limit.

\subsection{Effect of cross-layer hopping $t^d_{\perp13}$ \label{AppSec:t13} }
\begin{figure}[H]
\includegraphics[width=1\linewidth]{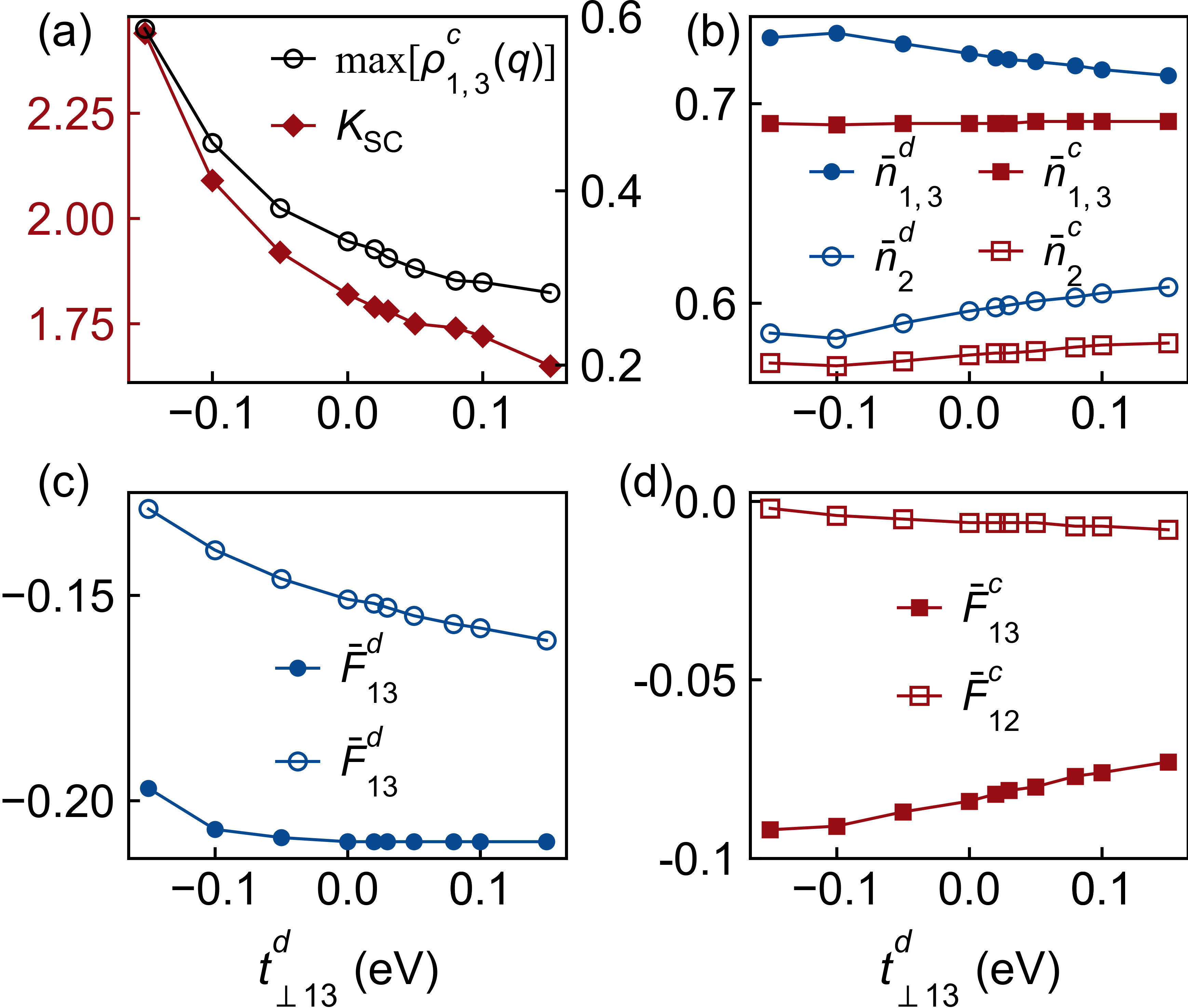} 
\caption{ Impact of cross-layer hopping $t^d_{\perp 13}$ on the two-orbital trilayer Hubbard model with fixed $U=3.5$ eV and $J_\mathrm{H}=1$ eV. (a) Luttinger parameter $K_\mathrm{SC}$ for cross-layer pairing $\Phi^{c\perp}_{1,3}$ (red) and the maximum value of $\rho^c_3(q)$ (black) in the outer layers of \XO orbital. (b) Average charge densities $\bar{n}^\alpha_\mu$ for both \ZO (circle) and \XO (square) orbitals. (c-d) Average interlayer and cross-layer spin correlation for (c) \ZO and (d) \XO orbitals. All presented data correspond to the $D\to\infty$ limit.
}
\label{FigA1}
\end{figure}

Figure~\ref{FigA1} shows the results with varying the cross-layer hopping, $t^d_{\perp 13}\in[-0.15,0.15]$~eV, in the two-orbital trilayer Hubbard model with fixed parameters $J_\mathrm{H}=1$ eV and $U=3.5$ eV. For the \ZO orbital, increasing $t^d_{\perp 13}$ leads to: (i) a redistribution of charge density ($n^d_{1,3}$ decreases while $n^d_2$ increases); (ii) nearly constant cross-layer spin correlation $\bar{F}^d_{13}$; and (iii) enhanced interlayer correlation $\bar{F}^d_{12}$, leading to an increased pairing frustration for the \ZO orbital. 

For the \XO orbital, $\bar{F}^c_{12}$ remains negligibly small and the cross-layer correlation $\bar{F}^c_{13}$ slightly weakens with $t^d_{\perp 13}$. 
Further more, both inner and outer layers display minor charge modulation, while the CDW order exhibits a marked suppression, as clearly demonstrated by the decreasing maximum in the charge structure factor $\rho^c_{1,3}(q)$ [Fig.~\ref{FigA1}(a)]. As a result, we observe a general trend where $K_\mathrm{SC}$ decreases monotonically as $t^d_{\perp 13}$ increases from -0.15 to 0.15 eV, demonstrating that both the magnitude and sign of $t^d_{\perp 13}$ influence the SC pairing. It also shows that despite its small magnitude ($t^d_{\perp 13}=0.078$ eV) in \LNO, this hopping parameter enhances the cross-layer pairing in the \XO orbital compared to the cases with zero or positive $t^d_{\perp 13}$. This finding highlights the previously overlooked role of $t^d_{\perp 13}$ in stabilizing SC order in \LNO.
 
Interestingly, this finding bears striking resemblance to observations in overdoped trilayer cuprate Bi$_2$Sr$_2$Ca$_2$Cu$_3$O$_{10+\delta}$,  where small cross-layer hopping amplitudes play a crucial role in maintaining high $T_c$ by modulating the charge distribution across Fermi surfaces and enhancing cross-layer SC pairing~\cite{Luo2023Electronic}. The cross-layer hopping may also contribute to the enhanced SC gap in the outer CuO$_2$ layer of trilayer cuprate (Hg, Re)Ba$_2$Ca$_2$Cu$_3$O$_{8+\delta}$ \cite{Horio2025Enhanced}. The observation of analogous behavior across diverse superconductors compels systematic theoretical investigation into the fundamental principles governing multilayer superconductivity.

\end{appendix}

\end{multicols}
\end{document}